\documentclass[a4paper]{article}
\usepackage{graphicx}
\usepackage{onecolpceurws}
\usepackage{booktabs}
\usepackage{graphicx}
\usepackage{caption}
\usepackage{subcaption}
\usepackage{amssymb}
\usepackage{color}

\title{A data-driven workflow for predicting horizontal well production using vertical well logs}

\author{
Jorge Guevara, Matthias Kormaksson,  Bianca Zadrozny  \\ IBM Research, Brazil\\ jorgegd, matkorm , biancaz@br.ibm.com \\
\and
\\ Ligang Lu, John Tolle, Tyler Croft, Mingqi Wu, Jan  Limbeck, Detlef Hohl \\ Shell Inc.\\
          { ligang.lu, John.Tolle, Tyler.Croft, Mingqi.Wu, J.Limbeck, detlef.hohl@shell.com}
}

\institution{}

\begin{document}
\maketitle

\begin{abstract}

In recent work, data-driven sweet spotting technique for shale plays previously explored with vertical wells has been proposed. Here, we extend this technique to multiple formations and formalize a general data-driven workflow to  facilitate feature extraction from vertical well logs and predictive modeling of horizontal well production. We also develop an experimental framework that facilitates model selection and validation in a realistic drilling scenario. We present some experimental results using this methodology in a field with 90 vertical wells and 98 horizontal wells,  showing that it can achieve better results in terms of predictive ability than kriging of known production values.

\end{abstract}
\vskip 32pt

\section{Introduction}

In recent years, interest in unconventional resource exploration has grown substantially, particularly in North America, due to horizontal drilling and hydraulic fracturing techniques. However, these new techniques come at a cost, which poses many operational challenges. For example, with drilling costs at an all time high, choosing the right locations for new wells is a crucial issue. In this scenario, identifying so called ``sweet spots'' with high potential for oil and gas is of great importance. Currently, the industry is in a state of ``trial-and-error'', with only immature research results available regarding the physical characteristics of sweet spots. This opens up a great opportunity to explore data-driven approaches to effectively learn to characterize sweet spots. To this end, there is a huge amount of available data that the industry has been collecting over several decades.

In recent work, a data-driven sweet spotting technique for shale plays previously explored with vertical wells has been proposed~\cite{kormaksson}. The technique involves three steps: 1) automatically extract features from vertical well log curves, within a single shale formation, using functional Principal Component Analysis (fPCA), 2) interpolate the extracted features from vertical well locations to horizontal well locations, and 3) build predictive models that relate interpolated features with horizontal well production. The method was tested using well log data from 2020 vertical wells and production data from 702 horizontal wells in a single field. 

Here, we extend this previous work to multiple formations and formalize a general data-driven workflow that involves a series of steps to generate normalized data frames that facilitate feature extraction from vertical well logs and predictive modeling of horizontal well production. We also develop an experimental framework that facilitates model selection and validation in a realistic drilling scenario. This method is applicable in both large and small sample size settings. We finally show some experimental results using this methodology in another field with 90 vertical wells and 98 horizontal wells.

\section{Methodology}

\begin{figure} 
    \centering
    \includegraphics[width=\textwidth]{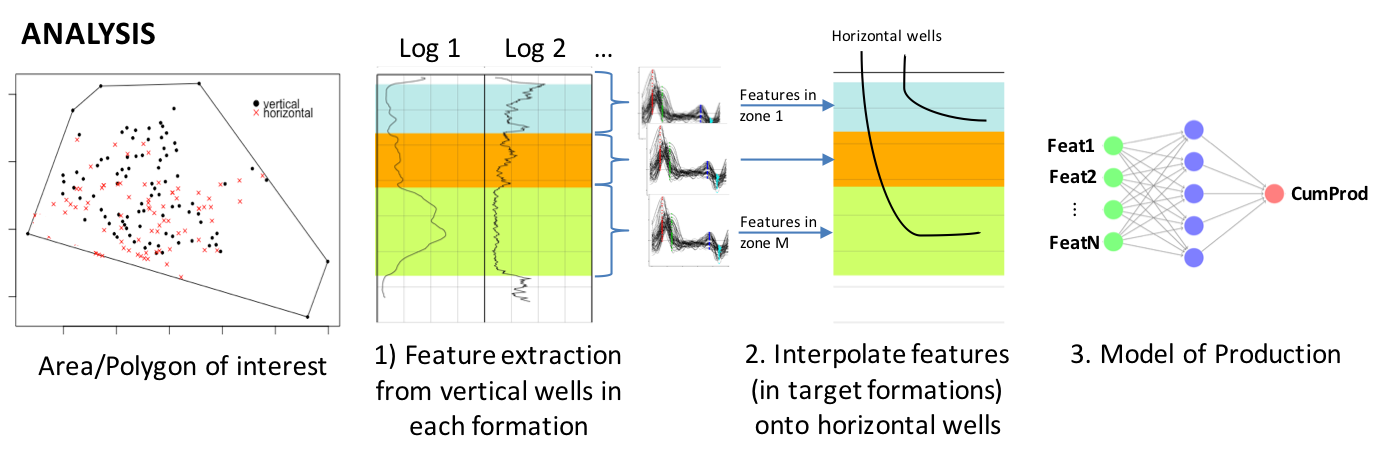} \caption{Analysis.} \label{fig:analysis}
\end{figure}

Our workflow is divided into several phases. Firstly, we perform data pre-processing, which is depicted in Figure~\ref{fig:preprocessing} and involves normalizing vertical and horizontal data sources to standardized data frames that may be used seamlessly in all downstream analyses. The data pre-processing is followed by an analysis workflow, which is depicted in Figure~\ref{fig:analysis} and consists of three main steps: 1) feature extraction from the (standardized) vertical well logs data frames, 2) interpolation of the extracted features onto horizontal well locations and incorporation of the features into the (standardized) production data frame, 3) predictive model building for horizontal well production with the aim to find sweet spots. 

\subsection{Data Pre-processing}

\begin{figure} 
    \centering
    \includegraphics[width=\textwidth]{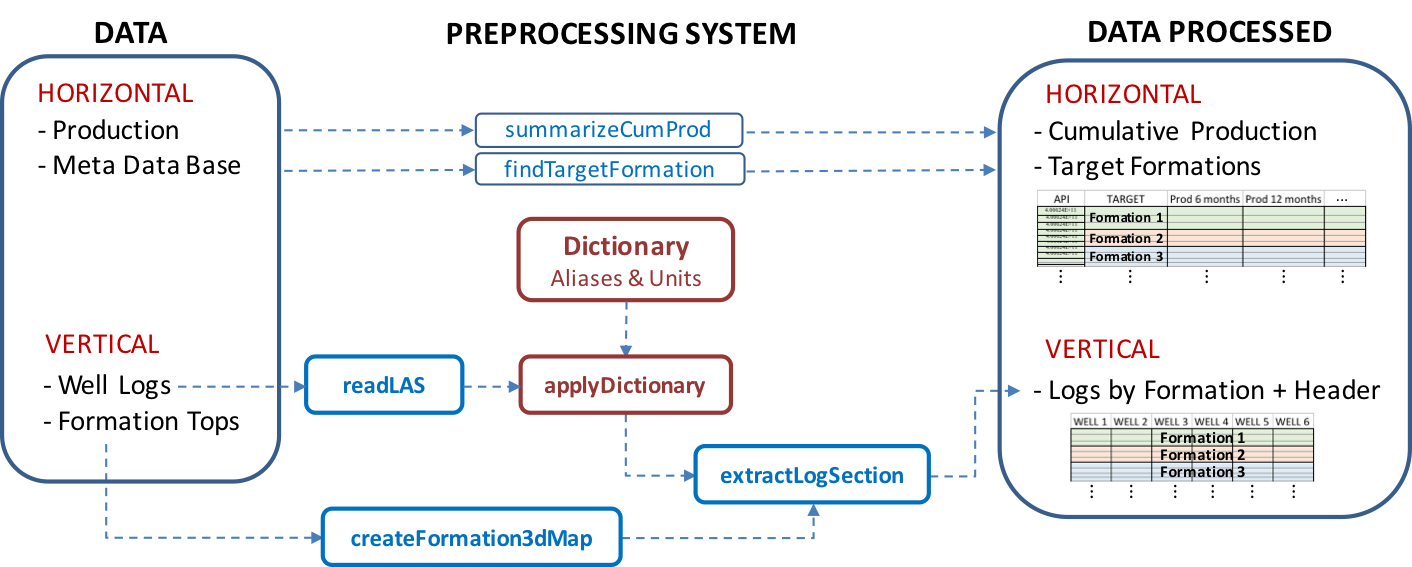} \caption{Pre-processing.} \label{fig:preprocessing}
\end{figure}
    
The data pre-processing system is demonstrated in Figure \ref{fig:preprocessing}. It receives as input a set of files that stores the horizontal and vertical data sources. The horizontal well data includes daily production for all the horizontal wells under study and a meta data base that contains information about the target formations. The target formation is defined as the principal formation where the horizontal section of a given well lands during drilling. 

The procedure ``summarizeCumProd'' takes as input the daily oil and gas production values and produces a cumulative production value for a chosen period of time (e.g. 6 months, 12 months, or 18 months). Cumulative production is calculated by summing the daily production values from the beginning of well production until the desired number of months later. The cumulative production will be the target variable for our predictive models. Wells that have produced for less time than the desired number of months receive a cumulative production value of ``NA'' (missing). Usually, we would like to choose the number of months reasonably large, but not too large so that we don't eliminate many wells from our analysis. 

The procedure ``findTargetFormation'' simply searches the meta data base for the target formations of the wells under study. Once the two pre-processing functions have been applied to the horizontal data then the data is stored in a data frame that contains the columns: ``API'', ``TARGET FORMATION'', ``Cum\_6month\_oil\_Prod'', ``Cum\_6month\_gas\_Prod'', ``Cum\_12month\_oil\_Prod'', etc. Each row of the data frame corresponds to a unique horizontal producing well in the area/polygon of study. This data frame is the data structure that will be accessed for all downstream analyses and we refer to it as the ``cumulative production data frame''. Whenever new features are generated for the horizontal wells they can be stored as new columns appended to this data frame, e.g. surface\_X, surface\_Y coordinates of the horizontal wells (from meta data base), or functional principal components (extracted from vertical well logs), see section \ref{sec:featureextraction}.

The vertical well data include well log files (.las) and files containing the formation tops of the vertical wells in the area/polygon of study. Formation tops of a given vertical well are stored as pairs of values (Formation\_Name, Depth) that determine the name of the given formation and the depth at which it starts. If for a given vertical well we have the top of each target formation along with the top of the next formation below, then we can determine the relevant vertical well log sections within all target formations. The function ``readLAS'' essentially reads the well logs contained in the .las files corresponding to vertical wells inside the area/polygon of interest. After reading the .las file of a given vertical well we store in memory the quantitative log values in a data frame whose columns are the depth values and the underlying log properties. 

In order for column names to be consistent across vertical wells (in .las files) we apply a function called ``applyDictionary''. This function will scan through well log (column) names and substitute each log name with a unique alias that represents the unique identifier of the corresponding well log property. For example, if three .las files have logs named ``GammaRay'', ``Gamma'', and ``GR'', respectively, then they may be substituted by the unique alias name ``GR''. The underlying dictionary is maintained by geophysics experts and may be automated in part by accessing online alias sources, such as Crain's Petrophysical Handbook. 

The ``createFormation3dMap'' procedure reads in the formation tops (and bottoms) of all target formations and infers the depths of target formation tops and bottoms at all vertical wells. In case of incomplete data (e.g. missing formation depth at a given well) we apply spatial interpolation techniques to infer approximate depths. 

The procedure ``extractLogSection'' is applied to each vertical well log data frame (as obtained from ``readLAS'') and uses the inferred formation 3D map (as obtained from ``createFormation3dMap'') to extract the relevant well log sections within all the target formations of interest. The ``extractLogSection'' procedure also re-samples (in depth) each well log section across all vertical wells so that a single data frame may be formed for each well log property of interest. More specifically, assume that for a given formation, each well $i$ contains $n_i$ depth values between the formation top and bottom, $i=1,\dots,N$. Since formation thicknesses vary in sizes across the different vertical wells (and thus the number of depth values differ), we cannot store the raw well log section (across all the vertical wells) in a single data frame. However, if we choose a constant number $n$ (e.g. $n = \sum_i n_i/N$) and we re-sample/interpolate the well log section of each well at $n$ equally spaced depth values between top and bottom, then we may store the formation's well log sections (across all the vertical wells) in a single data frame. Each formation can thus be stored as a sub-matrix whose columns correspond to the vertical wells and whose rows correspond to comparable depth values (across wells). The first and last row of the given sub-matrix would correspond to the formation top and bottom, respectively. The intermittent rows would correspond to a sequence of equally spaced depth values between top and bottom. The above essentially corresponds to a uniform depth normalization of each formation and results in a data frame that may be easily accessed and manipulated in all downstream analyses, see e.g. section \ref{sec:featureextraction}. The data frame that is obtained by stacking these sub-matrices across all target formations we refer to as ``standardized well log data frame'' and we note that we create a separate data frame for each well log property of interest.

\subsection{Feature Extraction per Formation} \label{sec:featureextraction} 
From the standardized (vertical) well log data frames we may now extract features (within each formation) that we wish to include in our predictive modeling efforts (see step 1 of the analysis workflow in Figure~\ref{fig:analysis}). These features may include simple summary statistics of the well logs, such as mean, variance, and maximum/minimum peak per formation. We extend the approach of~\cite{kormaksson} and calculate functional principal components within each formation independently. Note that the standardized well log data frames facilitate greatly the calculation of the principal components. For each formation we simply select the appropriate sub-matrix corresponding to that formation and then provide it as direct input into the appropriate functions of the R-package ``fda'' (see~\cite{Ramsay}), which we used for calculation of functional principal components. We calculated functional principal components of primary curves from a petrophysical perspective, which included Density, Gamma Ray, Limestone, Neutro Porosity, Deep and Shallow Resistivity, Photoelectric Factor, Medium Resistivity, Compressional and Shear Slowness.  

\subsection{Interpolation}
Once the features have been extracted from the vertical well log sections within each formation, we interpolate those features onto the coordinates of the horizontal production wells. However, since for a given horizontal well it has one and only one target formation, we only consider features extracted from vertical well log sections inside the underlying target formation. More specifically, for each horizontal well we gather the features of the vertical well log sections of the corresponding target formation and then interpolate onto the (surface) coordinates of the horizontal well. Once the features have been interpolated/inferred for all horizontal wells, the features may be appended as new columns to the cumulative production data frame, which can then be used directly for building predictive models.

\subsection{Predictive Modeling}\label{sec:predictiveModeling}
This phase explores the relationship between the interpolated fPCA features and the cumulative production for oil or gas at horizontal wells by means of predictive modeling. 
To this end, we have established an experimental framework for ranking, selecting and validating machine learning models, see section~\ref{sec:experiments}. Our framework uses a data set defined by the fPCA values (features) and cumulative production for gas or oil (predicted values) for all the horizontal wells from the \emph{cumulative production data frame}. Once a training set is defined, we select the most predictive features (fPCA values) using feature selection techniques. After that, we implement and rank several machine learning models in order to determine the best models. Once the best models have been established we perform an external validation to measure predictive power of the models in a realistic scenario and estimate the degree of association between the selected fPCA features and the cumulative production.

\section{Experimental Framework} \label{sec:experiments}

In order to select and validate predictive models, we propose an experimental framework that implements several machine learning models, ranks them in terms of predictive power and then externally validates performance of best models in a realistic drilling scenario.

\subsection{Model Selection} \label{subsec:modelselection}
In order to identify the best predictive models for a given training data set, we perform a benchmark experiment as suggested in \cite{Hothorn}. The goal of this experiment is to rank a set of candidate models in terms of the root mean squared error (RMSE). Some models may have built-in feature selection capabilities and for those models we enable feature selection using their own internal algorithms. For models that lack such capabilities we perform feature selection using the Elastic Net method \cite{zou2005regularization}. In order to avoid over-fitting, i.e., selecting features correlated to singularities on data, we select the subset of features with smallest cross validation error. After that, for each model, we estimate the distribution of the RMSE error using the re-sample technique described in \cite{Hothorn}. In our setting we use $K$-fold cross validation with $B$ repeats which results in an RMSE error distributed across $K \cdot B$ re-samples. The numbers $K$ and $B$ may be chosen in function of the underlying experimental data set (e.g. in our experiments in subsection \ref{subsec:experimentalresults} we chose $K=10$ and $B=3$). If a model has hyper-parameters to be optimized, then the distribution of the RMSE error is chosen by selecting the re-samples corresponding to the hyper-parameters with smallest cross validation error.
Finally, we select the top best models for the given training data in terms of the smallest median value of the distribution of the RMSE values. 

\subsection{Nested Leave-One-Out Validation} \label{subsec:LOO}
We perform an external validation in order to assess the performance of the selected best models from the benchmark experiment\footnote{We do not  use the RMSE values of the benchmark experiment because they could be optimistic in error estimation \cite{Cawley}.}. 
Through this external validation we further
``simulate'' the realistic scenario of predicting the cumulative production of a new (unseen) well given only information from previous existing wells. 
To this end, the feature and model selection steps of the previous sub-section are done within an external nested Leave-One-Out (LOO) loop (i.e. the validation procedure is external to the feature and model selection steps).
We proceed as follows:
we divide our complete data set of size $N$ into $N$ leave-one-well-out subsets of size $N-1$.
For each subset of size $N-1$ 
we performe feature and model selection using the methods described in the previous sub-section. 
Finally, the cumulative production is predicted for each of the omitted wells in the external LOO loop.

\subsection{Experimental Results} \label{subsec:experimentalresults}
We applied the experimental framework on two datasets both containing the same predictors but different predicted variables depending on whether the goal was to predict cumulative production of oil or gas. We denote those datasets as $D_{oil}$ and $D_{gas}$. The dataset $D_{oil}$ consisted of 88 non-missing observations (corresponding to 88 wells) that were chosen from 98 horizontal wells at some specific area of interest after selecting only those wells with twelve months of cumulative production. 
The same procedure was applied to the dataset $D_{gas}$, which consisted of 86 non-missing observations.
The complete candidate predictor set for both $D_{oil}$ and $D_{gas}$ was defined by the first 10 (interpolated) functional principal components of each of the primary curves specified in section \ref{sec:featureextraction}.

We implemented several models and feature selection methods from the Caret library \cite{kuhn2015caret} (which consists of 64 models). After that, we chose the top three best models in terms of the root mean squared error ($RMSE$) using the technqiues of sub-section \ref{subsec:modelselection}. In our setting we used $10$-fold cross validation with $3$ repeats which resulted in a RMSE error distributed across $30$ re-samples. Once top models were selected we conducted the nested leave-one-out cross validation (from sub-section \ref{subsec:LOO}) on those models in order to approximate the true generalization error as suggested in \cite{Cawley,ambroise2002selection}.

\begin{figure}
    \centering
    \begin{subfigure}[b]{0.65\textwidth}
        \includegraphics[width=\textwidth]{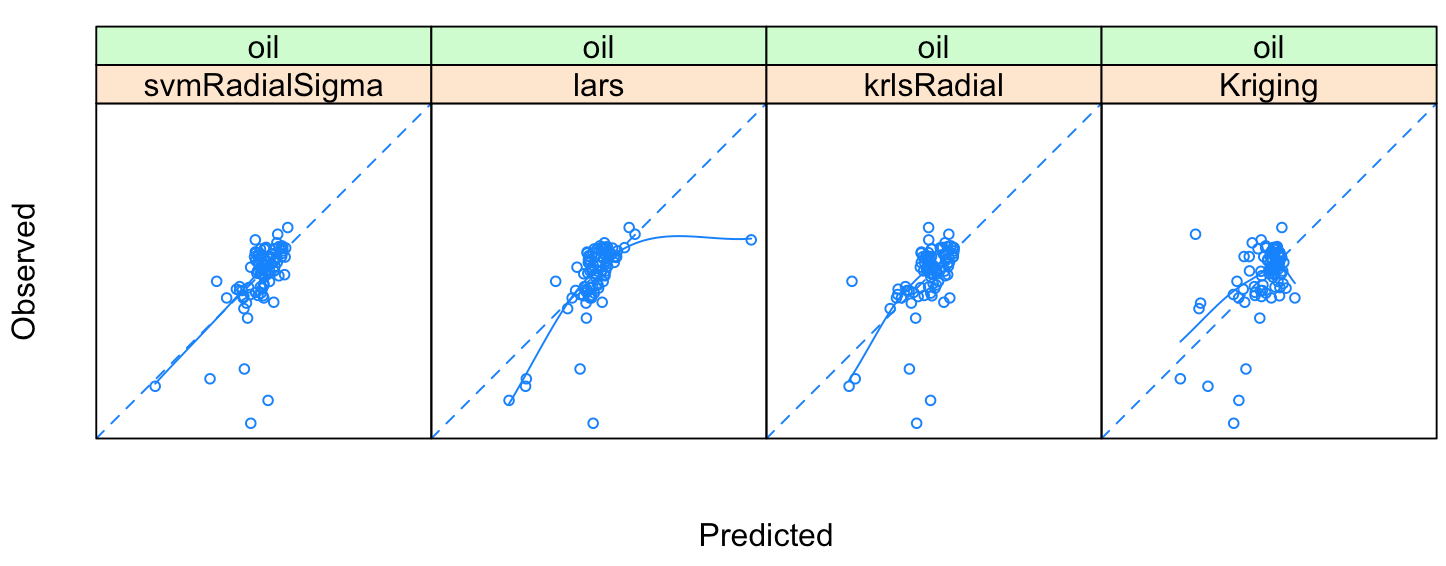}
    \end{subfigure}
    ~ 
    \begin{subfigure}[b]{0.65\textwidth}
        \includegraphics[width=\textwidth]{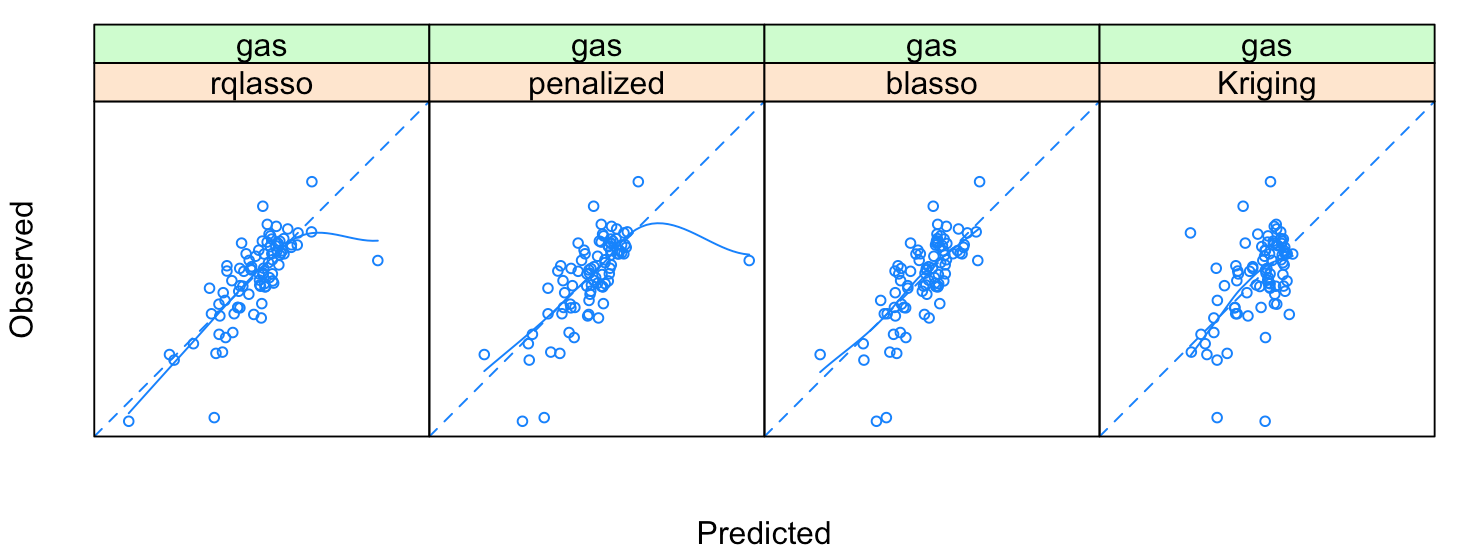}
    \end{subfigure}
    \caption{Predicted vs. observed values for oil production (above) and gas production (below) }\label{fig:ObsvsPred}
\end{figure}

Figure \ref{fig:ObsvsPred} shows the observed vs the predicted values obtained from the top three machine learning models. The figure also shows the observed vs predicted values using kriging on the raw horizontal well production.
Table \ref{Table:results} shows the LOO RMSE error,  the feature selection method  used and the Pearson correlation coefficient  for the top  regression models and kriging for oil and gas. We note that ``svmRadialSigma'' and ``rqlasso'' gave the best results for oil and gas in terms of RMSE values, respectively, and both outperformed traditional kriging.

\begin{table}[!htbp]
\caption{Results for the top 3 machine learning methods for oil and gas compared to kriging}
\begin{center}
\begin{tabular}{llll}
\toprule
Oil \\
\midrule
Method &Feature Selection& RMSE & Pearson Correlation \\
\midrule
svmRadialSigma& Elastic Net& $0.723$& $0.599$\\
lars &  built-in & $0.724$ &$0.607$ \\
krlsRadial & Elastic Net&$0.737$ &$0.573$\\
\midrule
kriging& horizontal production &$0.765$&$0.530$ \\
\bottomrule
\toprule
Gas && &\\
\midrule
Method &Feature Selection& RMSE & Pearson Correlation \\
\midrule
rqlasso&built-in &$0.522$&$0.767$\\
blasso &built-in &$0.546$&$0.742$ \\
penalized&built-in &$0.605$&$0.679$\\
\midrule
kriging&horizontal production&$0.704$ &$0.507$ \\
\bottomrule
\end{tabular}
\end{center}
\label{Table:results}
\end{table}


\section{Conclusions and Future Work}

The workflow we presented in this paper is a systematic methodology for using and comparing machine learning methods to predict production in unconventional plays using well logs and production data from previous explorations, taking into account multiple formations. Our experimental results show that this workflow can outperform more conventional techniques, such as kriging, in particular for the case of gas production prediction.

As future work, we would like to add to our workflow the capability of extracting and integrating features from horizontal well logs, in additional to features from vertical well logs. We also expect to be able to add well completion parameters in the models, which we expect will be very important in terms for improving the accuracy of the production predictions.


\begin{thebibliography}{Com79}

\bibitem[Amb02]{ambroise2002selection}
{Christophe Ambroise} {and} {Geoffrey~J McLachlan}.
\newblock {Selection bias in gene extraction on the basis of
  microarray gene-expression data}.
\newblock {\em Proceedings of the national academy of sciences\/} {99}, 10
  (2002), 6562--6566.
\newblock

\bibitem[Caw10]{Cawley}
{Gavin~C Cawley} {and} {Nicola~LC Talbot}.
\newblock {On over-fitting in model selection and subsequent
  selection bias in performance evaluation}.
\newblock {\em Journal of Machine Learning Research\/} {11}, Jul (2010),
  2079--2107.
\newblock

\bibitem[Hot05]{Hothorn}
{Torsten Hothorn}, {Friedrich Leisch}, {Achim Zeileis}, {and} {Kurt Hornik}.
\newblock {The design and analysis of benchmark experiments}.
\newblock {\em Journal of Computational and Graphical Statistics\/} {14}, 3
  (2005), 675--699.
\newblock

\bibitem[Kor15]{kormaksson}
{M.~Kormaksson}, {M.~Vieira}, {and} {B.~Zadrozny}
\newblock{A data driven method
for sweet spot identification in shale plays using well log data.}
\newblock{\em In SPE Digital
Energy Conference and Exhibition. (2015)} 
\newblock{Society of Petroleum Engineers.}

\bibitem[Kuh15] {kuhn2015caret}
{Max Kuhn}.
\newblock {Caret: classification and regression training}.
\newblock {\em Astrophysics Source Code Library\/}  {1} (2015), 05003.
\newblock


\bibitem[Ram09]{Ramsay}
{J. O. Ramsay}, {G. Hooker}, {and}
{S. Graves.} \newblock{Functional Data Analysis with R and
Matlab, (2009)}, \newblock{\em Springer}.

\bibitem[Zou05]{zou2005regularization}
{Hui Zou} {and} {Trevor Hastie}. 
\newblock {Regularization and variable selection via the
  elastic net}.
\newblock {\em Journal of the Royal Statistical Society: Series B (Statistical
  Methodology)\/} {67}, 2 (2005), 301--320.
\newblock


\end{thebibliography}

\end{document}